\documentclass[final,5p,times,twocolumn]{elsarticle}
\usepackage{lineno}

\usepackage{graphicx}
\usepackage{amssymb}

\def\etal{{\sl et al.}}                 

\journal{Nuclear Instruments and Methods A}

\begin{document}

\begin{frontmatter}

\title{Beam Test Studies of 3D Pixel Sensors Irradiated Non-Uniformly \\ 
for the ATLAS Forward Physics Detector}

\author[a]{S. Grinstein\corref{cor}\fnref{fn}}
\author[b]{M. Baselga}
\author[c]{M. Boscardin}
\author[d]{M. Christophersen}
\author[e]{C. Da Via}
\author[f]{G.-F. Dalla Betta}
\author[g]{G. Darbo}
\author[h]{V. Fadeyev}
\author[b]{C. Fleta}
\author[f]{C. Gemme}
\author[i]{P. Grenier}
\author[a]{A. Jimenez} 
\author[a]{I. Lopez}
\author[a]{A. Micelli} 
\author[g]{C. Nellist}
\author[j]{S. Parker}
\author[b]{G. Pellegrini}
\author[d]{B. Phlips}
\author[k]{D.-L. Pohl}
\author[h]{H. F.-W. Sadrozinski}
\author[l]{P. Sicho}
\author[a]{S. Tsiskaridze}           
\cortext[cor]{Corresponding author}
\fntext[fn]{Email-address: sgrinstein@ifae.es}

\address[a]{ICREA and Institut de F\'isica d'Altes Energies (IFAE), Barcelona, Spain}
\address[b]{Centro Nacional de Microelectronica, CNM-IMB (CSIC), Barcelona , Spain}
\address[c]{Fondazione Bruno Kessler, FBK-CMM, Trento, Italy}
\address[d]{U.S. Naval Research Laboratory, Washington, USA}
\address[e]{School of Physics and Astronomy, University of Manchester, Manchester, United Kingdom}
\address[f]{Universita degli Studi di Trento and INFN, Trento, Italy}
\address[g]{INFN Sezione di Genova, Genova, Italy}
\address[h]{Santa Cruz Institute for Particle Physics, University of California, Santa Cruz, USA}
\address[i]{SLAC National Accelerator Laboratory, Menlo Park, USA}
\address[j]{University of Hawaii, c/o Lawrence Berkeley Laboratory, Berkeley, USA}
\address[k]{University of Bonn, Bonn, Germany}
\address[l]{Institute of Physics ASCR v.v.i., Prague, Czech Republic}


\thispagestyle{empty}

\begin{abstract}
Pixel detectors with cylindrical electrodes that penetrate the silicon
substrate (so called 3D detectors) offer advantages over standard planar
sensors in terms of radiation hardness, since the electrode distance is
decoupled from the bulk thickness. 
In recent years significant progress has been made in the development of 3D
sensors, which culminated in the sensor production for the ATLAS Insertable
B-Layer (IBL) upgrade carried out at CNM (Barcelona, Spain) and FBK (Trento,
Italy).  Based on this success, the ATLAS Forward Physics (AFP) experiment
has selected the 3D pixel sensor technology for the tracking detector.
The AFP project presents a new challenge due to the need for a reduced
dead area with respect to IBL, and the in-homogeneous nature of the 
radiation dose distribution in the sensor. Electrical characterization of the 
first AFP prototypes and beam test studies of 3D pixel devices 
irradiated non-uniformly
are presented in this paper.
\end{abstract}


\end{frontmatter}

\section{Introduction}

\label{sec:introduction}


ATLAS~\cite{atlas} intends to install a Forward Physics detector (AFP) in
order to identify diffracted protons at $\approx$210~m from the 
interaction point~\cite{phase-1-LoI} in 2018. 
The current AFP design foresees a high resolution
pixelated silicon tracking system combined with a timing detector for the
removal of pile up protons.
The AFP tracker unit will consist of an array of six pixel sensors placed
at 2-3~mm from the Large Hadron Collider (LHC) proton beam. The proximity to 
the beam is essential
for the AFP physics program as it directly increases the sensitivity of
the experiment~\cite{phase-1-LoI}. 
%
Thus, there are two critical requirements for the AFP pixel detector:
first, the active area of the detector has to be as close as possible
to the LHC beam, which means that the dead region of the sensor has to be
minimized. Second, the device has to be able to cope with a
very inhomogeneous radiation distribution. Preliminary estimations 
indicate that the side of the sensors close to the beam will have to sustain 
a fluence equivalent to $\approx 5\times 10^{15}$ 1~MeV 
neutrons\footnote{The non-ionising energy loss will be quoted as the equivalent damage 
of a fluence of 1~MeV neutrons: n$_{\rm eq}$/cm$^2$.},
while the opposite site is expected to receive several orders of magnitude
less radiation~\cite{afp_cite}.

Based on the successful performance of CNM~\cite{cnm}
and FBK~\cite{fbk} 3D sensors productions~\cite{3dtech} for the ATLAS Insertable B-Layer
(IBL)~\cite{ibl_module_paper}, the 3D technology was chosen for the
AFP tracking detector. 
However, the IBL sensors have a large inactive edge on the side which 
will be closest to the beam in the AFP configuration.
To meet the AFP requirements, different dicing techniques have been 
investigated to reduce the dead area of the IBL 3D sensors.

As silicon sensors are irradiated the space charge in the detector
bulk changes, eventually leading to an increase in the full
depletion voltage.
Since the radiation distribution of the
AFP silicon sensors will be highly non-uniform, 
there can be a scenario where the breakdown voltage of the non-irradiated 
zone is lower than the depletion voltage of the irradiated one.
%
This will degrade the device performance: a high bias voltage will 
increase the leakage current leading to noise (to a point where the
device may not longer be operated), while an intermediate 
voltage will reduce the amount of charge collected in the irradiated
area, degrading the hit reconstruction efficiency.
Thus sensors with high initial breakdown voltage will be ideal for AFP.

This paper introduces the AFP pixel module in Section~\ref{sec:afp_module}.
Section~\ref{sec:prototypes} presents the results of electrical tests performed 
on slim-edge prototypes. The non-uniform irradiation of 3D sensors is described
in Section~\ref{sec:irrad}. Finally, in Section~\ref{sec:tb}, the performance 
of the irradiated devices in beam tests is presented.

\section{The AFP Pixel Module}

\label{sec:afp_module}

The pixel readout electronics of AFP will be the FE-I4~\cite{fei4} chip,
which was developed for IBL. Built in the 130~nm CMOS process, the FE-I4
offers an increased tolerance to radiation with respect to the current ATLAS
pixel readout chip, the FE-I3~\cite{fei3}. 
The IBL chip features an array of $336\times 80$ pixels with a pixel size of 
$50\times250$~$\mu$m$^2$. The total size of the FE-I4 chip is 
$20.2\times 19.0$~mm$^2$ and the active fraction is 89\%. 
%
The sensors are DC coupled to the chip with negative
charge collection. Each readout channel contains an independent 
amplification stage with adjustable shaping, followed by
a discriminator with independently adjustable threshold. The
chip operates with a $40$~MHz externally supplied clock. 
The time over threshold (ToT) with $4$-bit resolution
together with the firing time are stored for a latency interval
until a trigger decision is taken.

The selected sensor for AFP is the IBL 3D sensor, since it provides
high radiation tolerance at low bias voltage~\cite{sgrinstein_nima}.
The IBL 3D sensor is 230~$\mu$m thick with the n- and p-type columns 
etched from the opposite sides of the p-type substrate. The pixel configuration 
consists of two n-type readout electrodes connected at the wafer surface 
along the 250~$\mu$m long pixel direction, surrounded by six p-type 
electrodes which are shared with the neighboring pixels.
The CNM 3D sensor design features $210$~$\mu$m long columns which are
isolated on the n$^+$ side with p-stop implants. The edge isolation is
accomplished with a combination of a n$^+$ 3D guard ring,  which is
grounded, and fences, which are at the bias voltage potential from the
ohmic side.

\begin{figure}[!t]
\centering
\includegraphics[width=7cm]{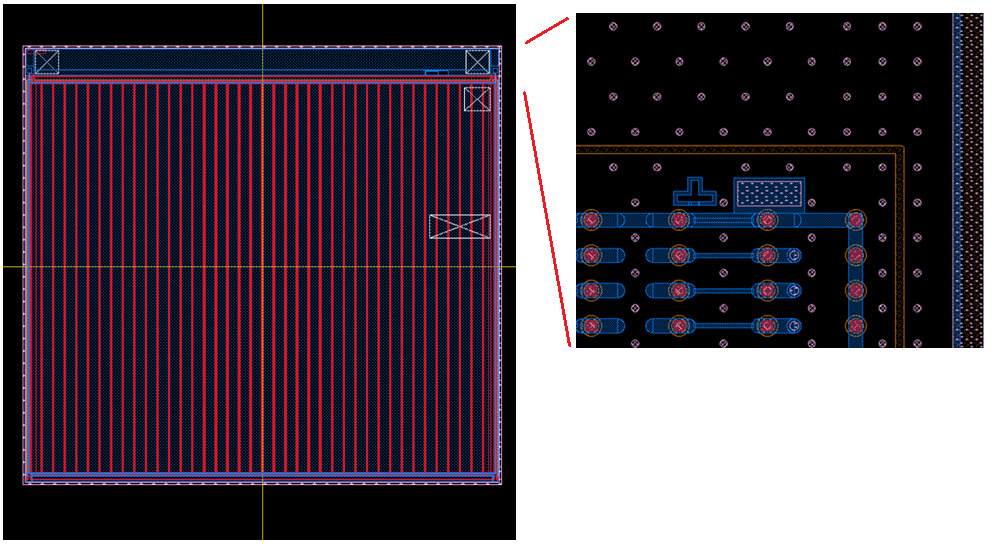}
\caption{Design mask of the CNM 3D sensor for IBL. The large dead area in the 
region opposite to the wirebonds is visible. A magnified view shows details of pixel 
geometry.}
\label{fig:design}
\end{figure}

%
The IBL sensor design has a large ($\approx$1mm) dead region in the side opposite to the 
wirebonds, see Fig~\ref{fig:design}. 
This dead region was added to accommodate the bias stab needed for the active
edge technology which was finally not incorporated into the IBL (the
side opposite to the wirebonds is not critical for the IBL, since the
sensors overlap in the $r-\phi$ direction). However, for the
AFP project, it is critical to substantially reduce the mentioned dead area,
since it will be closest to the beam.

In order to adapt the 3D IBL sensors for AFP, different techniques have been 
investigated.
Detectors were post-processed using the scribe-cleave-passivate
(SCP) technology to reduce the dead area.
Details of SCP method have been described in other publications~\cite{scp}. The 
method relies on passivating the sidewall with low defect density. 
A low defect density is obtained in a two-step process. First, the 
surface of the sensor is scribed at the desired edge location in the direction that 
coincides with one of the silicon crystal planes. Then 
the peripheral region is mechanically cleaved off. The resulting sidewall follows the crystal plane. 
The passivation step uses dielectric material that depends on the bulk type of 
the silicon. For p-type wafers used in this study, a dielectric with 
negative interface charge on the border with silicon is needed. This is accomplished 
by depositing an atomic layer (ALD) of Al$_2$O$_3$.
Standard diamond-saw cuts were also
investigated~\cite{fbk_marco_povoli_iword_2011}, but are not presented 
here.

\section{Electrical Tests of the First 3D Sensors Prototypes for AFP}

\label{sec:prototypes}

The SCP technology has been used to
reduce the inactive area of the IBL sensors for AFP. Since a limited number
of sensors were available at the time, the technique was first tested on 3D
sensors designed for the FE-I3 readout chip. 
The $230$~$\mu$m-thick FE-I3 3D sensors feature an array of $160\times 18$ pixels. 
Each $50\times 400$~$\mu$m$^2$ pixel is defined by  
three $210$~$\mu$m deep n-type readout 
columns, surrounded by eight p-type column electrodes.
Though the pixel
geometry is different, the results are expected to be relevant for 
the FE-I4 sensors, since the slim edge performance mostly depends on 
characteristics of the sensor periphery: sensor thickness, slim edge 
distance, and sidewall treatment. 

\begin{figure}[!t]
\centering
\includegraphics[width=7cm]{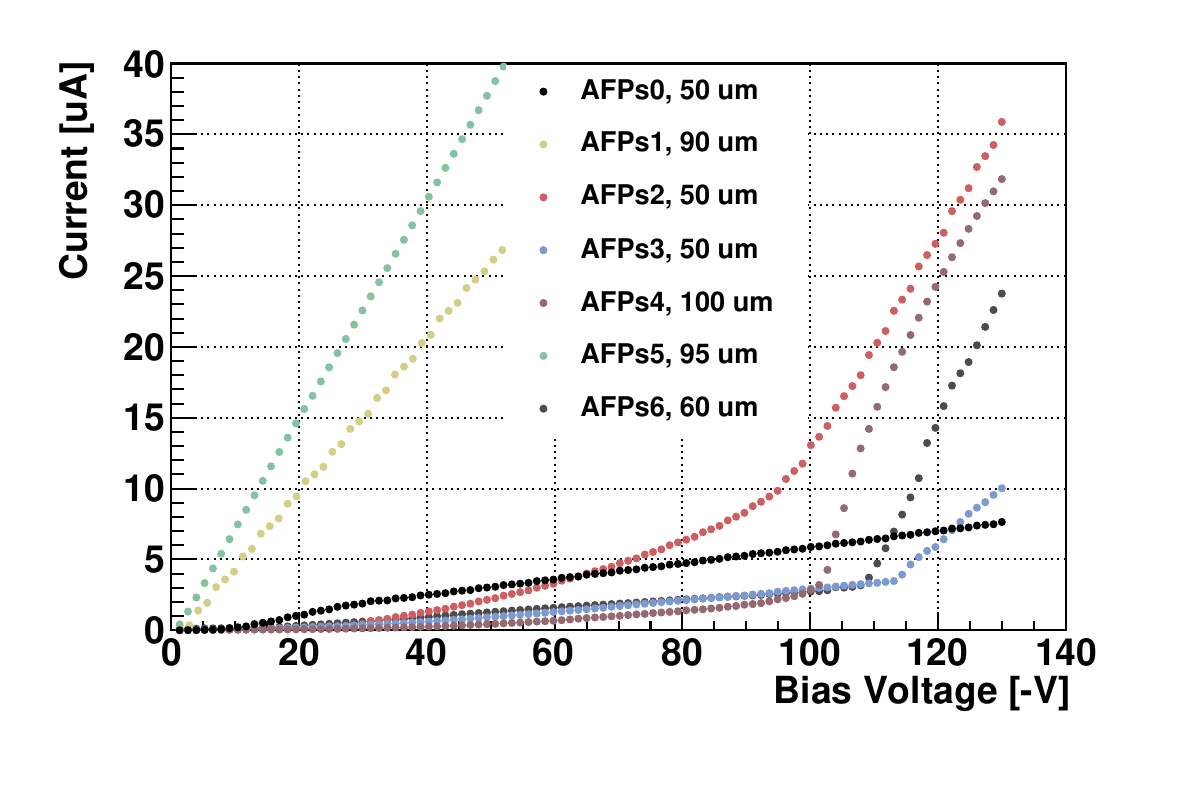}
\includegraphics[width=6.2cm]{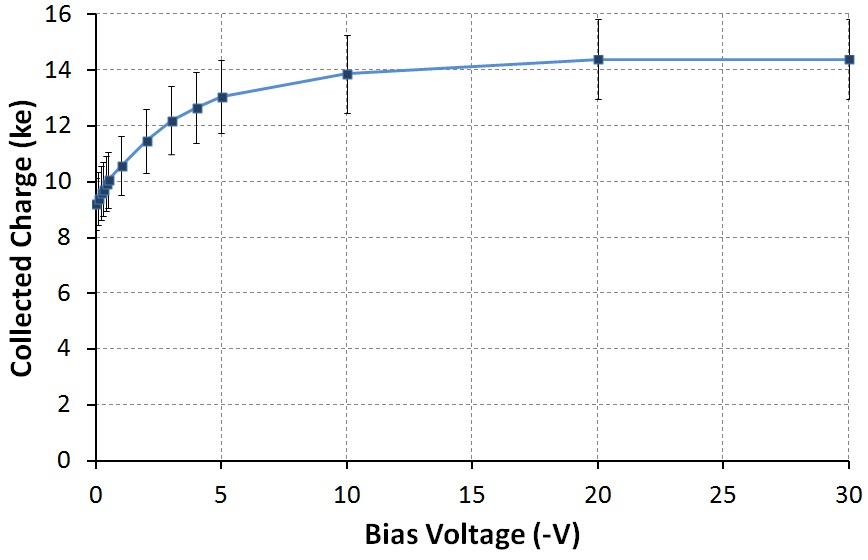}
\caption{Current versus bias voltage for CNM FE-I3 3D sensors with slim edges (top). 
The devices present low leakage currents and high breakdown voltage, with two exceptions
that have resistive behavior. 
The MPV of the Landau fit to the charge spectrum obtained 
with a Sr-90 source is shown below for the AFPs2 device (bottom). The error 
is dominated by the systematic uncertainty of the fit procedure.}
\label{fig:fei3s}
\end{figure}

Seven CNM 3D FE-I3 devices which presented breakdown voltages above $40$~V
were sent to NRL~\cite{nrl} to reduce the dead region by applying the SCP technology, from the
original $1$~mm to $50-100\,\mu$m. 
The devices
were returned to Barcelona to be bump-bonded to FE-I3 readout chips and
characterized. Figure~\ref{fig:fei3s} shows 
the leakage currents as a function of the bias voltage measured at room temperature.
Two assemblies show a resistive behavior. 
This may be caused by 
defects on the sidewall due to imperfect cleaving process.
%
The charge collection was verified in devices without resistive  
behavior using a 90-Sr source and an external trigger provided 
by a scintillator.
Figure~\ref{fig:fei3s} presents the most probable value (MPV) of the Landau fit to the 
charge distribution. Since the devices are 230~$\mu$m thick, the expected
charge at full depletion is about 15~k electrons, which is within the systematic 
uncertainties of the measurement above 10~V.



\section{Non-uniform Irradiation of 3D Sensors}

\label{sec:irrad}

One critical aspect of the pixel devices for the AFP forward detector is
the non-uniform nature of the radiation distribution across the sensor.
%
AFP pixel devices must be able to operate with high efficiency when only
a portion of the sensor is irradiated to $5\times 10^{15}$~n$_{\rm eq}$/cm$^2$.
This means that the less irradiated areas of the sensor have to sustain high
bias voltages keeping the leakage current at moderate values. Previous
experiences with FE-I4 CNM 3D devices indicated that the leakage current
has to be lower than 200~$\mu$A (front-end chip not powered) in order to
maintain low noise levels and ensure high hit reconstruction 
efficiency~\cite{sgrinstein_nima}.

\begin{figure}[!t]
\centering
\includegraphics[width=8cm]{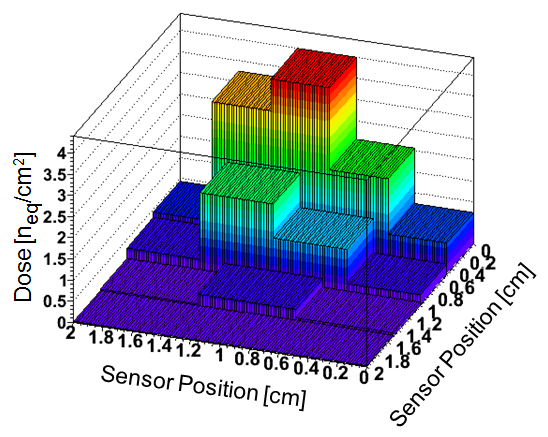}
\caption{Dose distribution on the CNM-57 sample~\cite{dosimetry}. A similar 
profile was obtained for CNM-83, but with a maximum dose of 
$9.4\times 10^{15}$~n$_{\rm eq}$/cm$^2$, see also Table~\ref{table:fei4_devices}.}
\label{fig:irrad}
\end{figure}

To study the effect of non-uniform irradiations, two FE-I4 IBL prototypes
(CNM-57 and CNM-83, without slim edges) were irradiated with protons 
non-uniformly at the IRRAD1 facility at CERN-PS~\cite{irrad1}. The irradiation levels are
listed in Table~\ref{table:fei4_devices}. Figure~\ref{fig:irrad} shows the
irradiation dose distribution on CNM-57, a similar profile was obtained
for CNM-83 but with a larger maximum dose.

\begin{figure}[!t]
\centering
\includegraphics[width=7cm]{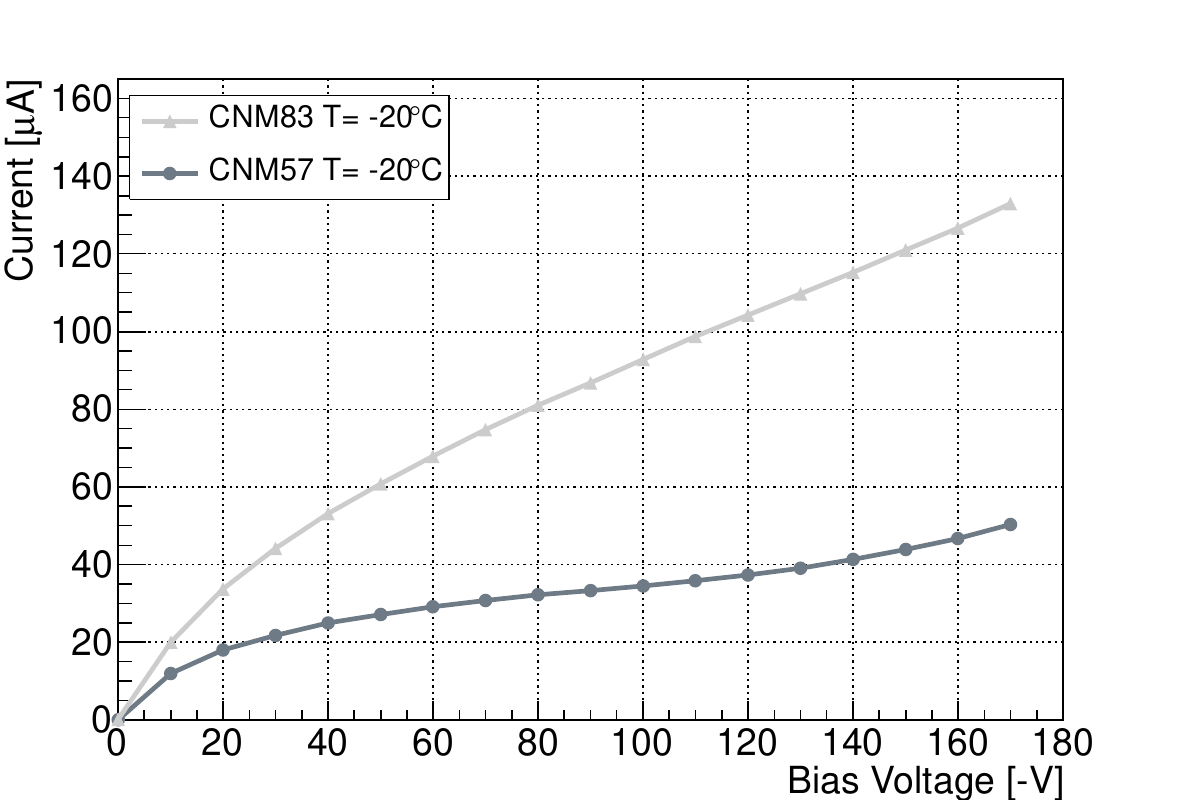}
\caption{Measurement of the leakage current as a function of the bias 
voltage for the two irradiated CNM devices at $-20^{\circ}$C with the
front-end chip not powered.}
\label{iv_fei4_irrad}
\end{figure}

The leakage currents measured as a function of the bias voltage for
CNM-57 and CNM-83 are shown in Fig.~\ref{iv_fei4_irrad}.  The measurements
were taken at $-20^{\circ}$~C with the front-end electronics not powered\footnote{The 
temperature was determined with a Pt-1000 temperature sensor 
on the module.}.
The device with early breakdown before irradiation shows a large leakage current at intermediate
bias voltages ($\approx$150~V), an indication that its performance may
not be optimal due to the noise induced by the dark current.
%

\begin{table}[!t]\footnotesize
\renewcommand{\arraystretch}{1.3}
\caption{Samples irradiated non-uniformly for AFP. The breakdown voltages
shown are before irradiation.
}
\label{table:fei4_devices}
\centering
\begin{tabular}{|c|c|c|c|}
\hline
Sample & Max. Dose                   & Pre-Irradiation &   Bias Voltage \\
       & ($10^{15}$~n$_{\rm eq}$/cm$^2$) & Breakdown & during Beam     \\
       &                                 & Voltage (v) & Tests (V)   \\
\hline
CNM-57  & 4.0      & 75                    & 130              \\
CNM-83  & 9.4      & 10                    & 130              \\ 
\hline
\end{tabular}
\end{table}
%

\begin{figure}[!t]
\centering
\includegraphics[width=8cm]{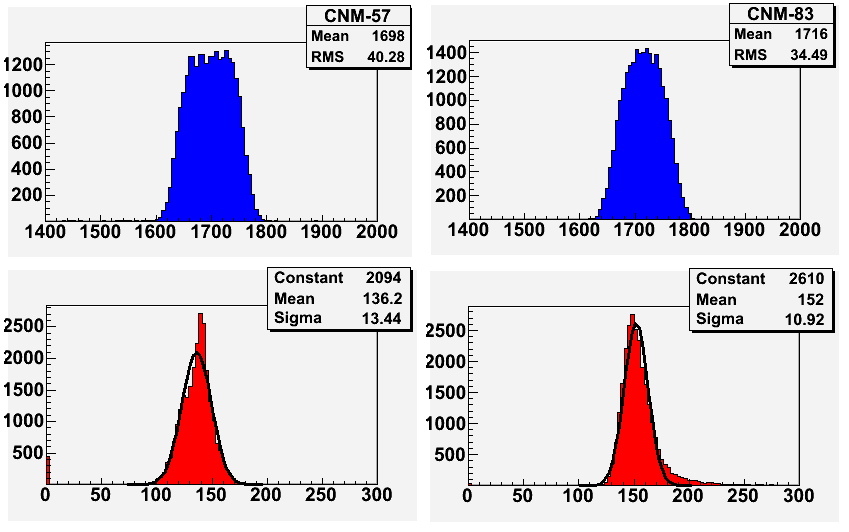}
\caption{Threshold (top) and noise (bottom) distribution of the 
FE-I4 CNM AFP prototype irradiated devices: CNM-57 (left) and CNM-83 (right).
}
\label{fig:thresholds}
\end{figure}

Before the performance of the devices is studied in beam tests, it is
necessary to verify that the front-end electronics was not damaged as a
result of the proton irradiation.
%
The operational threshold was set to $1700$ electrons based 
on previous experience with the IBL devices~\cite{ibl_module_paper}.
The threshold distributions
of the two irradiated devices, shown in Fig.~\ref{fig:thresholds}, 
present good uniformity, 
while the ENC (Equivalent Noise Charge) is below 200 electrons.

\section{Beam Test Studies of Non-uniformly Irradiated 3D Sensors}

\label{sec:tb}

Test beam studies of non-uniformly irradiated devices are essential to understand
the AFP pixel detector module performance. 
%
Critical parameters, such as hit efficiency and
position resolution, can only be determined at beam tests.
Both CNM-83 and CNM-57, together with an un-irradiated reference device, 
have been characterized using 120 GeV pions at the CERN SPS H6 beam line 
in August 2012.

Beam particle trajectories were reconstructed using the high resolution
EUDET telescope~\cite{eudet}. The telescope consists of six Mimosa tracking
planes, the readout data acquisition system and the trigger hardware, and
provides a $\approx3\mu$m track pointing resolution. The devices under
test, cooled to $-15^{\circ}$~C, were placed between the telescope planes. 
Data presented here were recorded at perpendicular incident angle. 
The hit efficiency is determined from extrapolated 
tracks on the devices, after track quality cuts have been applied. A
hit on the device under test is searched for in a $3\times3$ pixel
window around the track position.

\begin{figure}[!t]
\centering
\includegraphics[width=7cm]{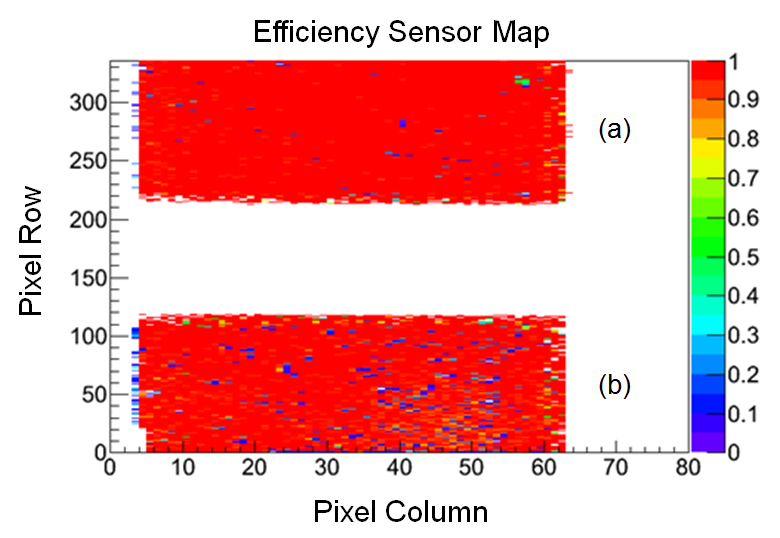}
\includegraphics[width=7cm]{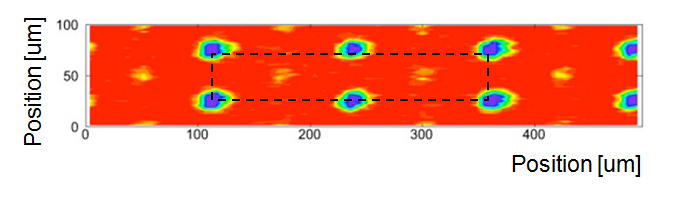}
\caption{Hit efficiency for CNM-57. The top plot shows the efficiency
across the sensor. The area labeled ``a'' (``b'') corresponds to the 
region which was less (more) irradiated (see Fig.~\ref{fig:irrad}). 
After dead and noise pixels are removed, the average efficiency for 
``a'' (``b'') is $99.2\%$ ($98.0\%$).
The efficiency for ``a'' folded
into a two by two pixel area is shown below.
}
\label{fig:cnm57_eff_sensor}
\end{figure}

Since the active area of the FE-I4 devices is larger than the Mimosa sensors 
of the telescope, separate sets of data were taken to cover the irradiated and 
non-irradiated regions of the sensors. Fig.~\ref{fig:cnm57_eff_sensor} shows the efficiency
map for sensor CNM-57. The non-irradiated region has an average efficiency of
$98.9\%$, while the efficiency for the irradiated side is $92.7\%$. If the 
dead and noise pixel cells (due to front-end issues) are removed, the efficiency
increases to $98.0\%$. 
In order to highlight the pixel structure, the corresponding efficiency as
function of the track hit position folded into a two by two cell is also shown
in Fig.~\ref{fig:cnm57_eff_sensor}. As the device is positioned perpendicular
to the particle beam the effect of the pass-through electrodes 
is evident. 
The efficiency for sensor CNM-83 on the irradiated side was much lower, about 
$60\%$, due to the large noise induced by the leakage current. 
This was expected from the electrical measurements done before the beam tests.
In fact, it was difficult to operate the device during the beam test because the
noise affected the data readout synchronization.

\section{Conclusions}

The AFP project presents a new challenge for pixel detectors: reduced
dead areas are needed to maximize the physics potential of the experiment,
and the sensors have to be able to sustain a highly non-uniform irradiation 
distribution. 
The scribe-cleave-passivate technology used to reduce the inactive edge of 
3D sensors showed promising results with FE-I3 devices. The first FE-I4 
prototypes have already been produced and will be tested shortly.
The first beam test studies of non-uniformly irradiated 3D pixel 
sensors have been presented. 
For devices with low initial breakdown, the performance is poor due to
the large leakage current caused by the bias voltage needed to deplete the
irradiated area.
Devices that show good electrical behavior before irradiation are able
to sustain the voltage needed to achieve excellent efficiency ($>98\%$
at perpendicular incidence) throughout the sensor.


\section*{Acknowledgment}

This work was partially funded by the MINECO, Spanish Government,
under the grant FPA2010-22060-C02-01/02, and the European Commission 
under the FP7 Research Infrastructures project AIDA, grant agreement 
no. 262025.
We would like to thank the Institute for Nanoscience (NSI) at the U.S. 
Naval Research Laboratory (NRL) and the NSI staff. The work done at NRL 
was sponsored by the Office of Naval Research (ONR). The work at SCIPP 
was supported by Department of Energy, grant DE-FG02-04ER41286.



%


\end{document}